\documentclass[12pt,reqno]{amsart}%
\usepackage{graphicx}
\usepackage{amscd}
\usepackage{amsmath}
\usepackage{epsfig}
\usepackage{amsfonts}
\usepackage{amssymb}%
\providecommand{\U}[1]{\protect\rule{.1in}{.1in}}
\numberwithin{equation}{section}
\providecommand{\U}[1]{\protect\rule{.1in}{.1in}}
\providecommand{\U}[1]{\protect\rule{.1in}{.1in}}
\allowdisplaybreaks

\theoremstyle{plain}

\begin{document}
\title[Degenerate Parametric Amplification]{Degenerate Parametric Amplification of Squeezed Photons: Explicit Solutions,
Statistics, Means and Variances}
\author{P. B. Acosta-Hum\'{a}nez }
\address{Department of Mathematics and Statistics, Universidad del Norte, KM 5 via
Puerto Colombia, Barranquilla, Colombia}
\email{pacostahumanez@uninorte.edu.co}
\author{S. I. Kryuchkov}
\address{School of Mathematical and Statistical Sciences, Arizona State University,
Tempe, AZ 85287--1804, U.S.A.}
\email{sergeykryuchkov@yahoo.com}
\author{E. Suazo}
\address{School of Mathematical and Statistical Sciences, Arizona State University,
Tempe, AZ 85287--1804, U.S.A \& School of Mathematical Sciences, University of
Puerto Rico, Mayaguez, Puerto Rico 00681-9000}
\email{erwin.suazo@upr.edu}
\author{S. K.~Suslov}
\address{School of Mathematical and Statistical Sciences, Arizona State University,
Tempe, AZ 85287--1804, U.S.A.}
\email{sks@asu.edu}
\urladdr{http://hahn.la.asu.edu/\symbol{126}suslov/index.html}
\date{April 12, 2015.}

\begin{abstract}
In the Schr\"{o}dinger picture, we find explicit solutions for two models of
degenerate parametric oscillators in the case of multi-parameter squeezed
input photons. The corresponding photon statistics and Wigner's function are
also derived in coordinate representation. Their time evolution is
investigated in detail. The unitary transformation and an extension of the
squeeze/evolution operator are briefly discussed.

\end{abstract}
\maketitle



\section{Introduction}

A model of the non-degenerate parametric amplifier was introduced and studied
in detail in classical papers \cite{LoisellYarivSiegman61},
\cite{GordonLouisellWalker63}, \cite{MollowGlauberI}, \cite{MollowGlauberII}
(see also reviews \cite{Hillery09} and \cite{Yariv11} for a historical
perspective and textbooks/reviews \cite{BachorRalph04}, \cite{Klyshko88},
\cite{Klyshko94}, \cite{Klyshko96}, \cite{Klyshko11}, \cite{MandelWolf},
\cite{Scully:Zubairy97} and \cite{Walls:Milburn} for a standard paradigm in
quantum optics). In a nonlinear dielectric medium, one adds to the linear
susceptibility tensor the second and third terms in the power expansion of the
polarization in the electric field. This nonlinear polarization couples back
to the electric field and a subsequent field quantization results in
interaction Hamiltonians that are cubic in the field. Among typical nonlinear
effects of this kind are the optical parametric amplification, the
second-harmonic generation (both third-order), and the degenerate four-wave
mixing (fourth-order) (see \cite{MollowGlauberI}, \cite{BachorRalph04},
\cite{Hanamuraetal07}, \cite{Hillery09} and the references therein for more
details). Nonlinear media are essential for the generation of squeezed and
entangled states of light \cite{AdessoIlluminati07},
\cite{BraunsteinvanLoock05}, \cite{Breit:Schill:Mlyn97}, \cite{Giedketal01},
\cite{LvRay09}, \cite{Mairetal01}, \cite{MolinaTerrizaetal07}. The squeezed
states and the quantum non-demolition experiment are expected to be utilized
in the detection of gravitational waves \cite{Caves81}, \cite{Abadetal11},
\cite{Demkowiczetal13} and also to enhance the performance of optical
communication systems \cite{BachorRalph04}, \cite{Hanamuraetal07}.

The degenerate parametric amplifier was investigated in \cite{Taka65},
\cite{Mollow67}, \cite{Raiford70}, \cite{Keating71},
\cite{PrakashChandraVach74}, \cite{Raiford74}, \cite{Stoler74}, \cite{Yuen76},
\cite{Rowe77}, \cite{HilleryZubairy82}, \cite{WodkiewichZubairy83},
\cite{Angelow:Trifonov95}, \cite{Angelow98} in the so-called parametric
approximation, when the pump mode is treated classically. Quantum description
beyond this approximation can be found in \cite{Raiford70},
\cite{HilleryZubairy84}, \cite{CohenBraunst95}, \cite{Hillery09} (see also the
references therein). Various aspects of the corresponding photon statistics
and photon-counting were studied in \cite{Glauber63}, \cite{Stoler70},
\cite{Stoler71}, \cite{Mollow73}, \cite{Caves81}, \cite{Caves82},
\cite{CollettGardiner84}, \cite{VenkSatya85}, \cite{Agarwal87},
\cite{CollettLoundon87}, \cite{VourdasWeiner87}, \cite{AgarwalAdam88},
\cite{FernCollett88}, \cite{FearnLoudon89}, \cite{KimOlivKnight89},
\cite{Marian91}, \cite{Marian92}, \cite{MarianMarianI93},
\cite{MarianMarianII93}, \cite{Dod:Man:Man94} in detail. Connections with the
experimentally observed dynamical Casimir effect \cite{Man'koCasimir},
\cite{Dodonov10}, \cite{Wilsonetal11}, \cite{Lahetal11} are discussed in
\cite{Dod09}, \cite{DezaelLambrecht10}, \cite{Johanetal10},
\cite{FujiietalZeilinger11}, \cite{Nationetal12}. All these results are also
of interest to the theory of quantum noise and measurement (see, for example,
\cite{Clerketal10}, \cite{Cavesetal12}, \cite{ChiXie13} and the references
therein). Nowadays, advanced experimental techniques allow one to measure
photon correlation functions of input microwave signals \cite{Bozyigitetal10},
to do quantum tomography on itinerant microwave photons \cite{Eichleretal11a},
and to study squeezing of microwave fields \cite{Eichleretal11b},
\cite{Malletetal11} (see also \cite{Braggioetal13}, \cite{Galeazzietal13} for
experimental study of a single-mode thermal field using a microwave parametric amplifier).

In spite of the considerable literature on the degenerate parametric
amplifiers, the general case of multi-parameter squeezed input photons
(corresponding, say, to a cascade of nonlinear crystals), to the best of our
knowledge, has never been discussed. Traditionally, the interaction picture is
commonly used \cite{Mollow67}, \cite{Scully:Zubairy97}, \cite{Walls:Milburn}
even though the statistics is postulated in the Schr\"{o}dinger picture
\cite{Klyshko94}, \cite{Klyshko98}. In this article, we study the statistical
properties of output squeezed quanta in terms of explicit solutions of certain
Ermakov-type system introduced in \cite{Lan:Lop:Sus}. In particular, explicit
formulas for the mean number and variance of generated photons are found
together with the corresponding time-dependent photon statistics. We elaborate
on the dynamical aspects of this problem related to evolution of the
corresponding photon states in Fock's space. In order to achieve this goal, we
utilize a unified approach to generalized harmonic oscillators discussed in
detail in several recent publications \cite{Cor-Sot:Lop:Sua:Sus},
\cite{Me:Co:Su}, \cite{Cor-Sot:Sus}, \cite{CorSus11}, \cite{Lan:Lop:Sus},
\cite{SanSusVin}, \cite{Lop:Sus:VegaGroup}, \cite{KretalSus13} (see also
\cite{Malk:Man:Trif73}, \cite{Dod:Mal:Man75}, \cite{Dodonov:Man'koFIAN87},
\cite{Malkin:Man'ko79} for the classical accounts). A similar treatment may be
useful for the Josephson metamaterial dynamical Casimir effect
\cite{Johanetal10}, \cite{Nationetal12}, \cite{Wilsonetal11}, \cite{Lahetal11}%
, \cite{FujiietalZeilinger11} (see also \cite{Dod95}, \cite{Dod09},
\cite{Law94}, \cite{Braggioetal05}), for experimental recognition of squeezed
microwave photons \cite{Eichleretal11b}, \cite{Malletetal11}, and for study of
a single-mode thermal microwave field \cite{Braggioetal13},
\cite{Galeazzietal13}.

The paper is organized as follows. In sections~2 and 3, we describe two
exactly solvable models of optical degenerate parametric amplifiers. The
generalized Fock states are constructed in section~4. The mean and variance of
the number operator are evaluated in Schr\"{o}dinger's picture in section~5.
The eigenfunction expansions of the generalized harmonic states of light in
terms of the standard Fock ones are derived in section~6 in coordinate
representation. In section~7, the Wigner and Moyal functions of the
multi-parameter squeezed states are evaluated directly from the corresponding
wavefunctions and their time evolution is verified with the help of a computer
algebra system. A brief summary is provided in the end. Our explicit solutions
of the corresponding Ermakov-type systems are given in Appendix~A together
with the means and variances of position and momentum operators. A convenient
expansion for the single photon mode Hamiltonian is presented in Appendix~B,
solutions in interaction picture are briefly discussed in Appendix~C, and a
canonical transformation of creation and annihilation operators is derived in
Appendix~D. An attempt to collect relevant references is made.

\section{Degenerate Parametric Amplifiers}

In this article, we consider the quantization of radiation field in a variable
dielectric medium in the Schr\"{o}dinger picture, as outlined in
\cite{Heitler57}, \cite{Ber:Lif:Pit} (in vacuum), by using the method of
dynamical invariants originally developed in \cite{Dod:Mal:Man75},
\cite{Dodonov:Man'koFIAN87}, \cite{Malk:Man:Trif73}, \cite{Malkin:Man'ko79}
and recently revisited in \cite{Cor-Sot:Sua:SusInv}, \cite{Suslov10},
\cite{SanSusVin}. To that end, we follow the mathematical technique of the
field quantization for a variable quadratic system in an abstract
(Fock-)Hilbert space discussed in \cite{KretalSus13} and concentrate on a
single mode of the radiation field. In this picture, the time evolution of
degenerate parametric amplifier is governed by the time-dependent
Schr\"{o}dinger equation for the state vector $\left\vert \psi\left(
t\right)  \right\rangle :$
\begin{equation}
i\frac{d}{dt}\left\vert \psi\left(  t\right)  \right\rangle =\widehat{H}%
\left(  t\right)  \left\vert \psi\left(  t\right)  \right\rangle
\label{Schroedinger}%
\end{equation}
with a certain variable quadratic Hamiltonian considered in original
publications \cite{Mollow67}, \cite{Raiford70}, \cite{Stoler74}. In a more
general setting, the degenerate parametric amplification with time-dependent
amplitude and phase was discussed by Raiford \cite{Raiford74}. The
corresponding Hamiltonian, without damping and neglecting high-frequency
terms, has the form%
\begin{equation}
\widehat{H}\left(  t\right)  =\frac{\omega}{2}\left(  \widehat{a}%
\ \widehat{a}^{\dagger}+\widehat{a}^{\dagger}\ \widehat{a}\right)
-\frac{\lambda\left(  t\right)  }{2}\left(  e^{i\left(  2\omega t+\phi\left(
t\right)  \right)  }\ \left.  \widehat{a}\right.  ^{2}+e^{-i\left(  2\omega
t+\phi\left(  t\right)  \right)  }\left(  \widehat{a}^{\dagger}\right)
^{2}\right)  . \label{RaiHam}%
\end{equation}
In this model, the phenomenological coupling parameter $\lambda\left(
t\right)  ,$ which describes the strength of the interaction between the
quantized signal of frequency $\omega$ and the classical pump of frequency
$2\omega,$ and the pump phase $\phi\left(  t\right)  $ are in general
functions of time. (It includes the special case of the pump and signal being
off-resonance by a given amount $\epsilon,$ i.~e., the pump frequency being
$2\omega+\epsilon,$ by letting $\phi\left(  t\right)  =\epsilon t$ and
$\lambda\left(  t\right)  =\lambda,$ a constant \cite{Raiford74}.) One can use
the standard annihilation and creation operators for a given mode $\omega,$%
\begin{equation}
\widehat{a}=\frac{1}{\sqrt{2\omega}}\left(  \omega\widehat{q}+i\left.
\widehat{p}\right.  \right)  ,\qquad\widehat{a}^{\dagger}=\frac{1}%
{\sqrt{2\omega}}\left(  \omega\widehat{q}-i\left.  \widehat{p}\right.
\right)  ,\qquad\left[  \widehat{a},\ \widehat{a}^{\dagger}\right]  =1,
\label{aaspq}%
\end{equation}
where $\widehat{q}$ and $\widehat{p}$ are time-independent operators in an
abstract Hilbert space with the canonical commutation relation $\left[
\left.  \widehat{q}\right.  ,\left.  \widehat{p}\right.  \right]  =i$ (in the
units of $\hbar).$ Then%
\begin{align}
&  \widehat{H}\left(  t\right)  =\frac{1}{2}\left(  1+\frac{\lambda\left(
t\right)  }{\omega}\cos\left(  2\omega t+\phi\left(  t\right)  \right)
\right)  \left.  \widehat{p}\right.  ^{2}\label{RaiHamPX}\\
&  \qquad+\frac{\omega^{2}}{2}\left(  1-\frac{\lambda\left(  t\right)
}{\omega}\cos\left(  2\omega t+\phi\left(  t\right)  \right)  \right)  \left.
\widehat{q}\right.  ^{2}\nonumber\\
&  \qquad\quad+\frac{\lambda\left(  t\right)  }{2}\sin\left(  2\omega
t+\phi\left(  t\right)  \right)  \left(  \widehat{p}\ \widehat{q}%
+\widehat{q}\ \widehat{p}\right) \nonumber
\end{align}
and the corresponding characteristic equation (classical equation of motion
\cite{Cor-Sot:Sua:SusInv}, \cite{Lan:Lop:Sus}) takes the form \cite{CorSus11}:%
\begin{align}
&  \mu^{\prime\prime}+\frac{\lambda\sin\left(  2\omega t+\phi\right)  \left(
2\omega+\phi^{\prime}\right)  -\lambda^{\prime}\cos\left(  2\omega
t+\phi\right)  }{\omega+\lambda\cos\left(  2\omega t+\phi\right)  }\mu
^{\prime}\label{RaiCharMu}\\
&  \quad\ +\frac{\omega\left(  \omega^{2}-3\lambda^{2}\right)  -\lambda
\phi^{\prime}-\lambda\left(  \omega^{2}+\lambda^{2}+\omega\phi^{\prime
}\right)  \cos\left(  2\omega t+\phi\right)  -\lambda^{\prime}\omega
\sin\left(  2\omega t+\phi\right)  }{\omega+\lambda\cos\left(  2\omega
t+\phi\right)  }\mu=0,\nonumber
\end{align}
which can be thought of as an extension of Ince's equation \cite{Mag:Win},
\cite{Menn68}. Here, we present two explicit solutions of this model when
$\lambda^{\prime}=0$ and $\phi=0,$ $\pi/2$ as usually accepted in the
literature. A general case can be discussed in a similar fashion.

\section{Two Integrable Cases}

For the Hamiltonian (\ref{RaiHamPX}) with $\lambda=$constant and $\phi=0:$%
\begin{align}
\widehat{H}\left(  t\right)   &  =\frac{1}{2}\left(  1+\frac{\lambda}{\omega
}\cos2\omega t\right)  \left.  \widehat{p}\right.  ^{2}+\frac{\omega^{2}}%
{2}\left(  1-\frac{\lambda}{\omega}\cos2\omega t\right)  \left.
\widehat{q}\right.  ^{2}\label{HamI}\\
&  +\frac{\lambda}{2}\sin2\omega t\ \left(  \widehat{p}\ \widehat{q}%
+\widehat{q}\ \widehat{p}\right)  ,\nonumber
\end{align}
the corresponding Ince's equation \cite{CorSus11}%
\begin{align}
&  \left(  \omega+\lambda\cos2\omega t\right)  \mu^{\prime\prime}%
+2\lambda\omega\sin2\omega t\ \mu^{\prime}\label{CharI}\\
&  \quad+\left(  \omega\left(  \omega^{2}-3\lambda^{2}\right)  -\lambda\left(
\omega^{2}+\lambda^{2}\right)  \cos2\omega t\right)  \mu=0\nonumber
\end{align}
has the following standard solutions:%
\begin{align}
\mu_{0}\left(  t\right)   &  =\left(  \sinh\lambda t\ \cos\omega
t+\cosh\lambda t\ \sin\omega t\right)  /\omega,\label{Stand1}\\
\mu_{1}\left(  t\right)   &  =\cosh\lambda t\ \cos\omega t+\sinh\lambda
t\ \sin\omega t,\nonumber
\end{align}
which have been recently found with the aid of differential Galois theory
\cite{AcostaSuazo13}, in particular, by using techniques for solving the
one-dimensional stationary Schr{\"{o}}dinger equation such as algebrization
procedure and Kovacic algorithm \cite{ac}, \cite{acmowe}. The Wronskian is
given by $W\left(  \mu_{0},\mu_{1}\right)  =-1-\left(  \lambda/\omega\right)
\cos2\omega t.$ (Traditionally, Ince's equation was studied for the sake of
periodic solutions, which do not exist for the degenerate parametric
oscillators under consideration \cite{CorSus11}, \cite{Mag:Win}.)

In the second case, when $\phi=\pi/2$ and%
\begin{align}
\widehat{H}\left(  t\right)   &  =\frac{1}{2}\left(  1-\frac{\lambda}{\omega
}\sin2\omega t\right)  \left.  \widehat{p}\right.  ^{2}+\frac{\omega^{2}}%
{2}\left(  1+\frac{\lambda}{\omega}\sin2\omega t\right)  \left.
\widehat{q}\right.  ^{2}\label{HamII}\\
&  +\frac{\lambda}{2}\cos2\omega t\ \left(  \widehat{p}\ \widehat{q}%
+\widehat{q}\ \widehat{p}\right)  ,\nonumber
\end{align}
we find standard solutions of the corresponding characteristic equation:%
\begin{align}
&  \left(  \omega-\lambda\sin2\omega t\right)  \mu^{\prime\prime}%
+2\lambda\omega\cos2\omega t\ \mu^{\prime}\label{CharII}\\
&  \quad+\left(  \omega\left(  \omega^{2}-3\lambda^{2}\right)  +\lambda\left(
\omega^{2}+\lambda^{2}\right)  \sin2\omega t\right)  \mu=0\nonumber
\end{align}
in a similar fashion:%
\begin{align}
\mu_{0}\left(  t\right)   &  =\frac{1}{\omega}e^{-\lambda t}\sin\omega
t,\label{Stand2}\\
\mu_{1}\left(  t\right)   &  =e^{\lambda t}\cos\omega t-\frac{\lambda}{\omega
}e^{-\lambda t}\sin\omega t\nonumber
\end{align}
with the Wronskian $W\left(  \mu_{0},\mu_{1}\right)  =1-\left(  \lambda
/\omega\right)  \sin2\omega t.$

\section{Generalized Fock States for Multi-Parameter Squeezed Photons}

The linear dynamical invariants have the form \cite{SanSusVin} (see also
Theorem~1 of Ref.~\cite{KretalSus13} for an abstract setting, which is adapted
here)\footnote{It represents a general time-dependent Bogoliubov's
transformation of the creation and annihilation operators.}:%
\begin{align}
\widehat{b}\left(  t\right)   &  =\frac{e^{-2i\gamma\left(  t\right)  }}%
{\sqrt{2}}\left(  \beta\left(  t\right)  \widehat{q}+\varepsilon\left(
t\right)  +i\frac{\widehat{p}-2\alpha\left(  t\right)  \widehat{q}%
-\delta\left(  t\right)  }{\beta\left(  t\right)  }\right)  ,
\label{LinDynInv}\\
\widehat{b}^{\dagger}\left(  t\right)   &  =\frac{e^{2i\gamma\left(  t\right)
}}{\sqrt{2}}\left(  \beta\left(  t\right)  \widehat{q}+\varepsilon\left(
t\right)  -i\frac{\widehat{p}-2\alpha\left(  t\right)  \widehat{q}%
-\delta\left(  t\right)  }{\beta\left(  t\right)  }\right)  .\nonumber
\end{align}
The real-valued solution of the corresponding Ermakov-type system
\cite{Lan:Lop:Sus}, \cite{Lop:Sus:VegaGroup} (subject to arbitrary real-valued
initial data $\alpha\left(  0\right)  ,$ $\beta\left(  0\right)  \neq0,$
$\gamma\left(  0\right)  ,$ $\delta\left(  0\right)  ,$ $\varepsilon\left(
0\right)  ,$ $\kappa\left(  0\right)  ,$ which provides a natural
multi-parameter description of the squeezing at $t=0$ \cite{KrySusVega13}) is
given by%
\begin{align}
&  \alpha\left(  t\right)  =\alpha_{0}\left(  t\right)  -\beta_{0}^{2}\left(
t\right)  \frac{\alpha\left(  0\right)  +\gamma_{0}\left(  t\right)  }%
{\beta^{4}\left(  0\right)  +4\left(  \alpha\left(  0\right)  +\gamma
_{0}\left(  t\right)  \right)  ^{2}},\label{SolErmakovI}\\
&  \beta\left(  t\right)  =-\frac{\beta\left(  0\right)  \beta_{0}\left(
t\right)  }{\sqrt{\beta^{4}\left(  0\right)  +4\left(  \alpha\left(  0\right)
+\gamma_{0}\left(  t\right)  \right)  ^{2}}},\label{SolIBeta}\\
&  \gamma\left(  t\right)  =\gamma\left(  0\right)  -\frac{1}{2}\arctan
\frac{\beta^{2}\left(  0\right)  }{2\left(  \alpha\left(  0\right)
+\gamma_{0}\left(  t\right)  \right)  }, \label{SolIGamma}%
\end{align}
and%
\begin{align}
\delta\left(  t\right)   &  =-\beta_{0}\left(  t\right)  \frac{\varepsilon
\left(  0\right)  \beta^{3}\left(  0\right)  +2\left(  \alpha\left(  0\right)
+\gamma_{0}\left(  t\right)  \right)  \delta\left(  0\right)  }{\beta
^{4}\left(  0\right)  +4\left(  \alpha\left(  0\right)  +\gamma_{0}\left(
t\right)  \right)  ^{2}},\label{SolErmakovII}\\
\varepsilon\left(  t\right)   &  =\frac{2\varepsilon\left(  0\right)  \left(
\alpha\left(  0\right)  +\gamma_{0}\left(  t\right)  \right)  -\beta\left(
0\right)  \delta\left(  0\right)  }{\sqrt{\beta^{4}\left(  0\right)  +4\left(
\alpha\left(  0\right)  +\gamma_{0}\left(  t\right)  \right)  ^{2}}%
},\label{SolIIEpsilon}\\
\kappa\left(  t\right)   &  =\kappa\left(  0\right)  -\varepsilon\left(
0\right)  \beta^{3}\left(  0\right)  \frac{\delta\left(  0\right)  }{\beta
^{4}\left(  0\right)  +4\left(  \alpha\left(  0\right)  +\gamma_{0}\left(
t\right)  \right)  ^{2}}\label{SolIIKappa}\\
&  +\left(  \alpha\left(  0\right)  +\gamma_{0}\left(  t\right)  \right)
\frac{\varepsilon^{2}\left(  0\right)  \beta^{2}\left(  0\right)  -\delta
^{2}\left(  0\right)  }{\beta^{4}\left(  0\right)  +4\left(  \alpha\left(
0\right)  +\gamma_{0}\left(  t\right)  \right)  ^{2}}\nonumber
\end{align}
in terms of the fundamental solutions:
\begin{align}
\alpha_{0}(t)  &  =\frac{1}{4a(t)}\frac{\mu_{0}^{\prime}(t)}{\mu_{0}(t)}%
-\frac{d(t)}{2a(t)},\quad\beta_{0}(t)=-\frac{1}{\mu_{0}(t)}, \label{SolFundam}%
\\
\gamma_{0}(t)  &  =\frac{\mu_{1}(t)}{2\mu_{0}(t)}+\frac{d(0)}{2a(0)}.\nonumber
\end{align}
Here,%
\begin{equation}
a\left(  t\right)  =\left\{
\begin{array}
[c]{c}%
\left(  1+\left(  \lambda/\omega\right)  \cos2\omega t\right)  /2,\medskip\\
\left(  1-\left(  \lambda/\omega\right)  \sin2\omega t\right)  /2
\end{array}
\right.  \qquad d\left(  t\right)  =\left\{
\begin{array}
[c]{c}%
\left(  \lambda\sin2\omega t\right)  /2\medskip,\\
\left(  \lambda\cos2\omega t\right)  /2
\end{array}
\right.  \label{CoeffsI-II}%
\end{equation}
and%
\begin{equation}
\alpha_{0}(t)=\left\{
\begin{array}
[c]{c}%
\dfrac{\omega}{2}\medskip\dfrac{\cosh\lambda t\ \cos\omega t-\sinh\lambda
t\ \sin\omega t}{\cosh\lambda t\ \sin\omega t+\sinh\lambda t\ \cos\omega t},\\
\dfrac{\omega}{2}\cot\omega t
\end{array}
\right.  \label{AlphaI-II}%
\end{equation}%
\begin{equation}
\beta_{0}(t)=\left\{
\begin{array}
[c]{c}%
-\dfrac{\omega}{\cosh\lambda t\ \sin\omega t+\sinh\lambda t\ \cos\omega
t}\medskip,\\
-e^{\lambda t}\dfrac{\omega}{\sin\omega t}%
\end{array}
\right.  \label{BetaI-II}%
\end{equation}%
\begin{equation}
\gamma_{0}(t)=\left\{
\begin{array}
[c]{c}%
\dfrac{\omega}{2}\medskip\dfrac{\cosh\lambda t\ \cos\omega t+\sinh\lambda
t\ \sin\omega t}{\cosh\lambda t\ \sin\omega t+\sinh\lambda t\ \cos\omega t},\\
\dfrac{\omega}{2}e^{2\lambda t}\cot\omega t
\end{array}
\right.  \label{GammaI-II}%
\end{equation}
for the Hamiltonians (\ref{HamI}) and (\ref{HamII}), respectively. Equations
(\ref{AlphaI-II})--(\ref{GammaI-II}) define the corresponding Green's
functions (see \cite{AcostaSuazo13}, \cite{Cor-Sot:Lop:Sua:Sus},
\cite{CorSus11} and \cite{Suslov11} for more details; in the limit
$\omega\rightarrow0$ for the second Hamiltonian, we obtain an interesting
model of a strong coupling; the propagator in the interaction picture is
traditionally used in the physical literature \cite{Mollow67},
\cite{Scully:Zubairy97}, \cite{Walls:Milburn}\footnote{See also Appendix~C for
a brief summary.}). Explicit forms of solutions (\ref{SolErmakovI}%
)--(\ref{SolIIKappa}) for both Hamiltonians are presented in Appendix~A; see
(\ref{AlphaI})--(\ref{M(t)}) and (\ref{AlphaII})--(\ref{KappaII}), respectively.

The corresponding dynamical Fock states $\left\vert \psi_{n}(t)\right\rangle
,$ where the phase $\kappa\left(  t\right)  $ finally shows up, can be
obtained from now on in a standard fashion \cite{Fock32-2}, \cite{Fock34-3},
\cite{Ber:Lif:Pit} with the aid of our variable creation and annihilation
operators (\ref{LinDynInv}) (see also \cite{KobManin89}, in particular,
dialogues 8 and 9 and section 3.4, and (\ref{TimeSqueezeOper})). Under a
certain condition, they do satisfy the time-dependent Schr\"{o}dinger equation
(\ref{Schroedinger}) (see \cite{Fock28-2}, Lemma~2 of Ref.~\cite{KretalSus13}
and (\ref{TimeSqueezeOper}) for more details). Moreover, the wave functions of
degenerate parametric oscillators in coordinate representation are given by
equation (18) of \cite{Lan:Lop:Sus} in terms of our explicit solutions of the
Ermakov-type system; see also (\ref{WaveFunctionN}) below.

\section{Mean Photon Number, Variances and Identities}

The time-dependent variances \cite{KretalSus13}:%
\begin{align}
&  \sigma_{p}=\langle\left(  \left.  \Delta\widehat{p}\right.  \right)
^{2}\rangle=\left(  n+\frac{1}{2}\right)  \frac{4\alpha^{2}+\beta^{4}}%
{\beta^{2}},\quad\sigma_{q}=\langle\left(  \left.  \Delta\widehat{q}\right.
\right)  ^{2}\rangle=\left(  n+\frac{1}{2}\right)  \frac{1}{\beta^{2}%
},\label{Variences}\\
&  \quad\quad\qquad\sigma_{pq}=\frac{1}{2}\langle\Delta\widehat{p}%
\ \Delta\widehat{q}+\Delta\widehat{q}\ \Delta\widehat{p}\rangle=\left(
n+\frac{1}{2}\right)  \frac{2\alpha}{\beta^{2}}\nonumber
\end{align}
with an invariant \cite{Dod:Man:Man94}:%
\begin{equation}
\left\vert
\begin{array}
[c]{cc}%
\sigma_{p} & \sigma_{pq}\\
\sigma_{pq} & \sigma_{q}%
\end{array}
\right\vert =\sigma_{p}\sigma_{q}-\sigma_{pq}^{2}=\left(  n+\frac{1}%
{2}\right)  ^{2} \label{Dinvariant}%
\end{equation}
can be evaluated in terms of solutions of the Ermakov-type system for the
generalized Fock states (described in general by Lemma~2 of \cite{KretalSus13}%
; see also (\ref{TimeSqueezeOper})). Expressions (4.9)--(4.12) of
\cite{KretalSus13} provide a convenient generic form for all quadratic
operators under consideration.

As a result, directly in the Schr\"{o}dinger picture, the average number of
photons for these states $\left\vert \psi_{n}\left(  t\right)  \right\rangle
:$%
\begin{equation}
\left\langle \widehat{N}\right\rangle (t)=\left\langle \widehat{a}^{\dagger
}\ \widehat{a}\right\rangle =\left\langle \psi_{n}(t)\left\vert \frac
{1}{2\omega}\left(  \left.  \widehat{p}\right.  ^{2}+\omega^{2}\left.
\widehat{q}\right.  ^{2}\right)  -\frac{1}{2}\right\vert \psi_{n}%
(t)\right\rangle \label{AverageN}%
\end{equation}
is given by%
\begin{align}
\left\langle \widehat{N}\right\rangle  &  =\left(  n+\frac{1}{2}\right)
\frac{4\alpha^{2}+\beta^{4}+\omega^{2}}{2\omega\beta^{2}}-\frac{1}%
{2}\label{AverageNFinal}\\
&  +\frac{1}{2\omega}\left[  \left(  \delta-\frac{2\alpha\varepsilon}%
{\beta^{2}}\right)  ^{2}+\frac{\omega^{2}\varepsilon^{2}}{\beta^{2}}\right]
.\nonumber
\end{align}
For the reader's convenience, a useful expansion of the single photon mode
Hamiltonian, $\widehat{H}=\left(  \left.  \widehat{p}\right.  ^{2}+\omega
^{2}\left.  \widehat{q}\right.  ^{2}\right)  /2,$ in terms of our variable
creation and annihilation operators and solutions of the corresponding
Ermakov-type system is given in Appendix~B.

The general expression for the mean photon number can be significantly
simplified for the Hamiltonians (\ref{HamI}) and (\ref{HamII}) of the
degenerate parametric oscillators under consideration. Indeed, one can get in
terms of \textquotedblleft slow\textquotedblright\ variables only:%
\begin{align}
&  A\left(  t\right)  =\frac{4\alpha^{2}+\beta^{4}+\omega^{2}}{\beta^{2}%
}\label{Ainvariant}\\
&  =\left\{
\begin{array}
[c]{c}%
\medskip\dfrac{\left(  2\alpha\left(  0\right)  +\omega\right)  ^{2}+\beta
^{4}\left(  0\right)  \ }{2\beta^{2}\left(  0\right)  }e^{2\lambda t}%
\ \quad\qquad\\
+\dfrac{\left(  2\alpha\left(  0\right)  -\omega\right)  ^{2}+\beta^{4}\left(
0\right)  \ }{2\beta^{2}\left(  0\right)  }e^{-2\lambda t},\bigskip\\
\dfrac{\left(  4\alpha^{2}\left(  0\right)  +\beta^{4}\left(  0\right)
\right)  e^{-2\lambda t}\ +\omega^{2}\ e^{2\lambda t}}{\beta^{2}\left(
0\right)  }%
\end{array}
\right. \nonumber
\end{align}
and%
\begin{align}
&  B\left(  t\right)  =\left(  \delta-\frac{2\alpha\varepsilon}{\beta}\right)
^{2}+\frac{\omega^{2}\varepsilon^{2}}{\beta^{2}}\ \label{Binvariant}\\
&  =\left\{
\begin{array}
[c]{c}%
\dfrac{1}{2}\left(  \delta\left(  0\right)  -\dfrac{2\alpha\left(  0\right)
\varepsilon\left(  0\right)  }{\beta\left(  0\right)  }-\dfrac{\omega
\varepsilon\left(  0\right)  }{\beta\left(  0\right)  }\right)  ^{2}%
e^{2\lambda t}\medskip\quad\quad\\
+\dfrac{1}{2}\left(  \delta\left(  0\right)  -\dfrac{2\alpha\left(  0\right)
\varepsilon\left(  0\right)  }{\beta\left(  0\right)  }+\dfrac{\omega
\varepsilon\left(  0\right)  }{\beta\left(  0\right)  }\right)  ^{2}%
e^{-2\lambda t},\bigskip\\
\left(  \delta\left(  0\right)  -\dfrac{2\alpha\left(  0\right)
\varepsilon\left(  0\right)  }{\beta\left(  0\right)  }\right)  ^{2}%
e^{-2\lambda t}+\dfrac{\varepsilon^{2}\left(  0\right)  \omega^{2}}{\beta
^{2}\left(  0\right)  }\ e^{2\lambda t}%
\end{array}
\right. \nonumber
\end{align}
for (\ref{HamI}) and (\ref{HamII}), respectively, with the common invariants%
\begin{align}
\varepsilon^{2}+\frac{\delta^{2}}{\beta^{2}}\  &  =\varepsilon^{2}\left(
0\right)  +\frac{\delta^{2}\left(  0\right)  }{\beta^{2}\left(  0\right)
}=C,\label{Cinvariant}\\
\kappa-\frac{\delta\varepsilon}{2\beta}  &  =\kappa\left(  0\right)
-\frac{\delta\left(  0\right)  \varepsilon\left(  0\right)  }{2\beta\left(
0\right)  }=D. \label{Kinvariant}%
\end{align}
In compact form,%
\begin{equation}
\left\langle \widehat{N}\right\rangle =\left(  n+\frac{1}{2}\right)
\frac{A\left(  t\right)  }{2\omega}+\frac{B\left(  t\right)  }{2\omega}%
-\frac{1}{2} \label{NCompact}%
\end{equation}
(see also \cite{Breit:Schill:Mlyn97}, \cite{Dod:Man:Man94},
\cite{KrySusVega13} and the references therein).

For the initial coherent state, when $n=\alpha\left(  0\right)  =0$ and
$\beta^{2}\left(  0\right)  =\omega,$ one gets%
\begin{align}
&  \left\langle \widehat{N}\right\rangle =\sinh^{2}\lambda t\nonumber\\
&  +\left\{
\begin{array}
[c]{cc}%
\medskip\dfrac{1}{2}\left(  \varepsilon^{2}\left(  0\right)  +\dfrac
{\delta^{2}\left(  0\right)  }{\omega}\right)  \cosh2\lambda t-\dfrac
{\delta\left(  0\right)  \varepsilon\left(  0\right)  }{\sqrt{\omega}}%
\sinh2\lambda t & \text{for (\ref{HamI}),}\\
\dfrac{1}{2}\left(  \varepsilon^{2}\left(  0\right)  e^{2\lambda t}%
+\dfrac{\delta^{2}\left(  0\right)  }{\omega}e^{-2\lambda t}\right)
\qquad\qquad & \text{for (\ref{HamII}).}%
\end{array}
\right.  \label{Ncoherent}%
\end{align}
For the vacuum state, when $\delta\left(  0\right)  =\varepsilon\left(
0\right)  =0,$ we obtain a familiar expression from the theory of dynamical
Casimir effect and spontaneous parametric fluorence:%
\begin{equation}
\left\langle \widehat{N}\right\rangle =\sinh^{2}\lambda t \label{Nvacuum}%
\end{equation}
(see, for example, \cite{Dod95}, \cite{Dod09}, \cite{Dodonov10},
\cite{Klyshko88}, \cite{Yariv11} and the references therein).

The corresponding time-dependent variance:%
\begin{equation}
\text{Var\ }\widehat{N}=\left\langle \left(  \widehat{N}-\langle
\widehat{N}\rangle\right)  ^{2}\right\rangle =\left\langle \left(
\widehat{H}/\omega\right)  ^{2}\right\rangle -\left\langle \left(
\widehat{H}/\omega\right)  \right\rangle ^{2} \label{VarN}%
\end{equation}
can also be evaluated in terms of \textquotedblleft slow\textquotedblright%
\ variables only:%
\begin{align}
\text{Var\ }\widehat{N}  &  =\frac{A^{2}\left(  t\right)  -4\omega^{2}%
}{8\omega^{2}}\left[  \left(  n+\frac{1}{2}\right)  ^{2}+\frac{3}{4}\right]
\label{VarNFinal}\\
&  +\frac{A\left(  t\right)  B\left(  t\right)  -C}{\omega^{2}}\left(
n+\frac{1}{2}\right) \nonumber
\end{align}
with the help of expansion (\ref{HAB}). For the initial vacuum state, in
particular,
\begin{equation}
\text{Var\ }\widehat{N}=\frac{1}{2}\sinh^{2}2\lambda t. \label{VarNvacuum}%
\end{equation}
Moreover, the second-order intensity correlation function \cite{Glauber63},%
\begin{equation}
g^{\left(  2\right)  }=\frac{\left\langle \left(  \widehat{a}^{\dagger
}\right)  ^{2}\ \left.  \widehat{a}\right.  ^{2}\right\rangle }{\left\langle
\widehat{a}^{\dagger}\ \widehat{a}\right\rangle ^{2}}=1+\frac{\text{Var\ }%
\widehat{N}-\langle\widehat{N}\rangle}{\left.  \langle\widehat{N}%
\rangle\right.  ^{2}}, \label{SecondCorr}%
\end{equation}
is explicitly given in terms of (\ref{NCompact}) and (\ref{VarNFinal}) for the
multi-parameter squeezed number states $\left\vert \psi_{n}\left(  t\right)
\right\rangle $ (see also \cite{Agarwal87}, \cite{VourdasWeiner87},
\cite{KimOlivKnight89} and \cite{Marian91} for special cases). Explicit
expressions for averages of operators $\widehat{q}$ and $\widehat{p}$ and for
their time-dependent variances are given in Appendix~A.

For a complete quantum mechanical description of the nonclassical states of
light generated in the process of degenerate parametric amplification, the
corresponding photon statistics are required. In vacuum, an explicit
connection between the multi-parameter squeezed states and photon
distributions is found in our recent publication \cite{KrySusVega13} (in
coordinate representation, when $\widehat{q}=x$ and $\widehat{p}%
=-i\partial/\partial x$). It can be readily extended to the models of optical
degenerate parametric oscillators under consideration. One should replace
$\alpha\rightarrow\alpha/\omega,$ $\beta\rightarrow\beta/\sqrt{\omega},$
$\delta\rightarrow\delta/\sqrt{\omega}$ in (6.9)--(6.11) of
\cite{KrySusVega13} and a required modification of identities (6.11)--(6.20)
is as follows. A joint complex identity,%
\begin{equation}
\frac{\delta}{\beta}+i\varepsilon=\left(  \frac{\delta\left(  0\right)
}{\beta\left(  0\right)  }+i\varepsilon\left(  0\right)  \right)  e^{2i\gamma
}, \label{ComplexIdentityBoth}%
\end{equation}
holds for both Hamiltonians (\ref{HamI}) and (\ref{HamII}), which implies
(\ref{Cinvariant}). Moreover, by separating the \textquotedblleft
fast\textquotedblright,\ $\omega,$\ and \textquotedblleft
slow\textquotedblright,\ $\lambda,$ variables,%
\begin{equation}
\delta-\frac{2\alpha\varepsilon}{\beta}+i\frac{\omega\varepsilon}{\beta
}=e^{-i\omega t}\ \xi\left(  t\right)  , \label{ComplexIdentitiesII}%
\end{equation}
where%
\begin{equation}
\xi\left(  t\right)  =\left\{
\begin{array}
[c]{c}%
\left(  \delta\left(  0\right)  -\dfrac{2\alpha\left(  0\right)
\varepsilon\left(  0\right)  }{\beta\left(  0\right)  }\right)  \cosh\lambda
t-\dfrac{\omega\varepsilon\left(  0\right)  }{\beta\left(  0\right)  }%
\sinh\lambda t\medskip\quad\qquad\quad\\
+i\left(  \dfrac{\omega\varepsilon\left(  0\right)  }{\beta\left(  0\right)
}\cosh\lambda t-\left(  \delta\left(  0\right)  -\dfrac{2\alpha\left(
0\right)  \varepsilon\left(  0\right)  }{\beta\left(  0\right)  }\right)
\sinh\lambda t\right)  ,\bigskip\\
\left(  \delta\left(  0\right)  -\dfrac{2\alpha\left(  0\right)
\varepsilon\left(  0\right)  }{\beta\left(  0\right)  }\right)  e^{-\lambda
t}+i\dfrac{\omega\varepsilon\left(  0\right)  }{\beta\left(  0\right)
}e^{\lambda t}%
\end{array}
\right.  \label{KSI}%
\end{equation}
for the Hamiltonian (\ref{HamI}) and (\ref{HamII}), respectively. As a
by-product, we derive identities (\ref{Binvariant}).

In a similar fashion,%
\begin{equation}
\frac{\omega+\beta^{2}}{2}-i\alpha=\frac{1}{2}e^{i\omega t}\ \frac{\eta\left(
t\right)  }{z\left(  t\right)  },\qquad\frac{\omega-\beta^{2}}{2}%
+i\alpha=\frac{1}{2}e^{-i\omega t}\ \frac{\zeta\left(  t\right)  }{z\left(
t\right)  }, \label{ComplexIdentitiesIandII}%
\end{equation}
where%
\begin{equation}
z\left(  t\right)  =\left\{
\begin{array}
[c]{c}%
\dfrac{\omega\cos\omega t+\left(  2\alpha\left(  0\right)  +i\beta^{2}\left(
0\right)  \right)  \sin\omega t}{\omega}\cosh2\lambda t\medskip\quad\\
+\dfrac{\left(  2\alpha\left(  0\right)  +i\beta^{2}\left(  0\right)  \right)
\cos\omega t+\omega\sin\omega t}{\omega}\sinh2\lambda t,\bigskip\\
e^{2\lambda t}\omega\cos\omega t+2\alpha\left(  0\right)  \sin\omega
t+i\beta^{2}\left(  0\right)  \sin\omega t
\end{array}
\right.  \label{Zeta}%
\end{equation}
and%
\begin{equation}
\eta\left(  t\right)  =\left\{
\begin{array}
[c]{c}%
\left(  \omega+\beta^{2}\left(  0\right)  \right)  \cosh2\lambda
t\ +2\alpha\left(  0\right)  \sinh\lambda t\medskip\qquad\quad\\
-i\left(  2\alpha\left(  0\right)  \cosh2\lambda t\ +\medskip\left(
\omega-\beta^{2}\left(  0\right)  \right)  \sinh\lambda t\right)  ,\bigskip\\
-2i\alpha\left(  0\right)  +\beta^{2}\left(  0\right)  +\omega e^{2\lambda t}%
\end{array}
\right.  \label{Uta}%
\end{equation}
$\medskip$
\begin{equation}
\zeta\left(  t\right)  =\left\{
\begin{array}
[c]{c}%
\medskip\left(  \omega-\beta^{2}\left(  0\right)  \right)  \cosh2\lambda
t\ +2\alpha\left(  0\right)  \sinh\lambda t\qquad\quad\\
+i\left(  2\alpha\left(  0\right)  \cosh2\lambda t\ +\medskip\left(
\omega+\beta^{2}\left(  0\right)  \right)  \sinh\lambda t\right)  ,\bigskip\\
2i\alpha\left(  0\right)  -\beta^{2}\left(  0\right)  +\omega e^{2\lambda t}%
\end{array}
\right.  \label{Vta}%
\end{equation}
for the Hamiltonians (\ref{HamI}) and (\ref{HamII}), respectively.
(Computational details are left to the reader; we use Mathematica and Maple to
verify our calculations.) Vector $z\left(  t\right)  $ is related to the
complex parametrization found in \cite{KretalSus13}, \cite{KrySusVega13} (see
also \cite{Dod:Man79} and \cite{Har:Ben-Ar:Mann11} ). Having these
modifications in mind, in the next section, we will be able to derive variable
probability amplitudes and photon statistics in terms of hypergeometric series
which are similar to those in \cite{KrySusVega13} (see also \cite{Taka65},
\cite{VenkSatya85}, \cite{Marian91}, \cite{Marian92}).

\section{Time-Dependent Probability Amplitudes and Photon Statistics}

In coordinate representation, when $\widehat{q}=x$ and $\widehat{p}%
=-i\partial/\partial x,$ the wave functions of the optical degenerate
parametric oscillators under consideration take the form\footnote{A direct
Mathematica verification of these solutions is given by Christoph Koutschan.}%
\begin{equation}
\psi_{n}\left(  x,t\right)  =e^{i\left(  \alpha x^{2}+\delta x+\kappa\right)
+i\left(  2n+1\right)  \gamma}\sqrt{\frac{\beta}{2^{n}n!\sqrt{\pi}}%
}\ e^{-\left(  \beta x+\varepsilon\right)  ^{2}/2}\ H_{n}\left(  \beta
x+\varepsilon\right)  , \label{WaveFunctionN}%
\end{equation}
where $H_{n}\left(  x\right)  $ are the Hermite polynomials \cite{Ni:Su:Uv}
and explicit solutions of the corresponding Ermakov-type system are given by
(\ref{AlphaI})--(\ref{KappaI}) and (\ref{AlphaII})--(\ref{KappaII}) for the
Hamiltonians (\ref{HamI}) and (\ref{HamII}), respectively. (An important
special case $\lambda=0$ was originally investigated in \cite{Marhic78}; see
also \cite{Dod:Man79}, \cite{LopSusVegaHarm}, \cite{Lop:Sus:VegaGroup},
\cite{KrySusVega13} and the references therein. In paraxial optics, this
solution may be associated with a new model of periodic lens-like medium; see,
for example, \cite{Mah:Sua:Sus13}.) In terms of the stationary harmonic
oscillator wavefunctions,%
\begin{equation}
\Psi_{n}\left(  x\right)  =\left(  \frac{\omega}{\pi}\right)  ^{1/4}%
\frac{e^{-\omega x^{2}/2}}{\sqrt{2^{n}n!}}\ H_{n}\left(  x\sqrt{\omega
}\right)  , \label{HarmonicWaveFunctions}%
\end{equation}
the eigenfunction expansion has the form \cite{KrySusVega13}:%
\begin{equation}
\psi_{n}\left(  x,t\right)  =e^{i\left(  2n+1\right)  \gamma}\sqrt{\frac
{\beta}{\omega^{1/2}}}\sum_{m=0}^{\infty}C_{mn}\left(  t\right)  \ \Psi
_{m}\left(  x\right)  . \label{ExpansionGeneral}%
\end{equation}
The time-dependent coefficients can be found in terms of our solutions of the
Ermakov-type system as follows
\begin{align}
&  C_{mn}\left(  t\right)  =\sum_{k=0}^{\infty}M_{mk}\left(  \alpha
,\beta\right)  \ T_{kn}\left(  \varepsilon,\frac{\delta}{\beta},\kappa\right)
\label{ExpansionCoeffs}\\
&  =\sum_{k=0}^{\infty}T_{mk}\left(  \frac{\omega^{1/2}\varepsilon}{\beta
},\omega^{-1/2}\left(  \delta-\frac{2\alpha\varepsilon}{\beta}\right)
,\kappa-\frac{\alpha\varepsilon^{2}}{\beta^{2}}\right)  \ M_{kn}\left(
\alpha,\beta\right)  .\nonumber
\end{align}
Here,%
\begin{align}
&  T_{mn}\left(  A,B,\Gamma\right)  =i^{m-n}\frac{e^{i\left(  \Gamma
-AB/2\right)  }\ e^{-\nu/2}\ }{\sqrt{m!n!}}\ \left(  \frac{iA+B}{\sqrt{2}%
}\right)  ^{m}\left(  \frac{iA-B}{\sqrt{2}}\right)  ^{n}%
\label{HeisenbergGroupMatrix}\\
&  \qquad\times\ _{2}F_{0}\left(  -n,\ -m;\ -\frac{1}{\nu}\right)  ,\qquad
\nu=\left(  A^{2}+B^{2}\right)  /2\nonumber
\end{align}
and%
\begin{align}
&  M_{mn}\left(  \alpha,\beta\right)  =i^{n}\sqrt{\frac{2^{m+n}\omega}%
{m!n!\pi}}\ \Gamma\left(  \frac{m+n+1}{2}\right) \label{MartixSU(1,1)Hyper}\\
&  \times\ \frac{\left(  \dfrac{\omega-\beta^{2}}{2}+i\alpha\right)
^{m/2}\left(  \dfrac{\omega-\beta^{2}}{2}-i\alpha\right)  ^{n/2}}{\left(
\dfrac{\omega+\beta^{2}}{2}-i\alpha\right)  ^{\left(  m+n+1\right)  /2}%
}\nonumber\\
&  \times~_{2}F_{1}\left(
\begin{array}
[c]{c}%
-m,\quad-n\\
\dfrac{1}{2}\left(  1-m-n\right)
\end{array}
;\dfrac{1}{2}\left(  1\pm\frac{2i\beta\sqrt{\omega}}{\sqrt{4\alpha^{2}+\left(
\beta^{2}-\omega\right)  ^{2}}}\right)  \right)  .\nonumber
\end{align}
(We use the standard definition of generalized hypergeometric series and an
integral evaluated by Bailey \cite{Bailey48}; see \cite{KrySusVega13} for more details.)

These general expressions can be significantly simplified, once again, with
the help of identities found in the previous section. By separating the
\textquotedblleft fast\textquotedblright\ and \textquotedblleft
slow\textquotedblright\ variables, one gets%
\begin{align}
&  T_{mn}\left(  \varepsilon,\frac{\delta}{\beta},\kappa\right)  =e^{2i\left(
m-n\right)  \gamma}\ S_{mn},\qquad S_{mn}=\left.  T_{mn}\left(  \varepsilon
,\frac{\delta}{\beta},\kappa\right)  \right\vert _{t=0}\ ,\label{Mat1}\\
&  T_{mn}\left(  \frac{\omega^{1/2}\varepsilon}{\beta},\omega^{-1/2}\left(
\delta-\frac{2\alpha\varepsilon}{\beta}\right)  ,\kappa-\frac{\alpha
\varepsilon^{2}}{\beta^{2}}\right)  =e^{i\omega\left(  n-m\right)  t}%
\ R_{mn}\left(  t\right)  \label{Mat2}%
\end{align}
and%
\begin{equation}
M_{mn}\left(  \alpha,\beta\right)  =e^{-i\omega\left(  m+1/2\right)
t}e^{-i\left(  2n+1\right)  \gamma}\sqrt{\frac{\beta\left(  0\right)  }{\beta
}}\ N_{mn}\left(  t\right)  . \label{Mat3}%
\end{equation}
Here, by definition%
\begin{align}
&  R_{mn}\left(  t\right)  =\frac{i^{m+n}}{\sqrt{m!n!\left(  2\omega\right)
^{m+n}}}\ \ \xi^{m}\left(  \xi^{\ast}\right)  ^{n}\label{RT}\\
&  \quad\times\ e^{iD-B\left(  t\right)  /\left(  4\omega\right)  }\ _{2}%
F_{0}\left(  -n,\ -m;\ -\frac{2\omega}{B\left(  t\right)  }\right) \nonumber
\end{align}
and%
\begin{align}
&  N_{mn}\left(  t\right)  =i^{n}\sqrt{\frac{2^{m+n+1}\omega}{m!n!\pi}%
}\ \Gamma\left(  \frac{m+n+1}{2}\right)  \frac{\zeta^{m/2}\left(  \zeta^{\ast
}\right)  ^{n/2}}{\eta^{\left(  m+n+1\right)  /2}}\label{NT}\\
&  \quad\times~_{2}F_{1}\left(
\begin{array}
[c]{c}%
-m,\quad-n\\
\dfrac{1}{2}\left(  1-m-n\right)
\end{array}
;\dfrac{1}{2}\left(  1+2i\sqrt{\frac{\omega}{A\left(  t\right)  -2\omega}%
}\right)  \right)  .\nonumber
\end{align}
The asterisk denotes complex conjugation. (Quadratic transformation (6.8) of
\cite{KrySusVega13} can be used in order to complete our evaluation; see also
\cite{Taka65}, \cite{VenkSatya85}, \cite{Marian91}, \cite{Marian92},
\cite{MarianMarianI93} and Appendix~D. The special case $\lambda=0,$
$\omega=1$ corrects a typo in \cite{KrySusVega13}.)

As a result, we finally obtain%
\begin{equation}
\psi_{n}\left(  x,t\right)  =\sqrt{\frac{\beta\left(  0\right)  }{\omega
^{1/2}}}\sum_{m=0}^{\infty}c_{mn}\left(  t\right)  \ e^{-i\omega\left(
m+1/2\right)  t}\ \Psi_{m}\left(  x\right)  ,
\label{ExpansionFinalIndependent}%
\end{equation}
where the time-dependent probability amplitudes are given by%
\begin{equation}
c_{mn}\left(  t\right)  =\sum_{k=0}^{\infty}N_{mk}\left(  t\right)
\ S_{kn}=\sum_{k=0}^{\infty}R_{mk}\left(  t\right)  \ N_{kn}\left(  t\right)
\label{CoeffFinalDependent}%
\end{equation}
in terms of our \textquotedblleft slow\textquotedblright\ variables $\xi,$
$\eta,$ and $\zeta$ (\textquotedblleft adiabatic invariants\textquotedblright)
only for all real-valued initial data/constants of motion (of the
corresponding Ermakov-type system). Thus, the total probability amplitudes are
associated with the product of two infinite matrices related to the Poisson
and Pascal distributions; see \cite{KrySusVega13} for more details. The
quantities $\left\vert c_{mn}\left(  t\right)  \right\vert ^{2}$ explicitly
determine the corresponding variable photon statistics. In mathematical terms,
our expansion (\ref{ExpansionFinalIndependent}) gives a mapping between two
complete sets of vectors in $\mathcal{L}^{2}\left(
\mathbb{R}
\right)  ;$ namely, the transition matrix (\ref{CoeffFinalDependent}) between
the generalized harmonic, or \textquotedblleft squeezed\textquotedblright%
\ states, $\left\{  \psi_{n}\left(  x,t\right)  \right\}  _{n=0}^{\infty},$
and the standard Fock ones, $\left\{  e^{-i\omega\left(  m+1/2\right)
t}\ \Psi_{m}\left(  x\right)  \right\}  _{n=0}^{\infty},$ in coordinate
representation; the latter are usually being recorded by a photodetector in
clever quantum nonlinear optics experiments (see, for example,
\cite{Klyshko94}, \cite{Klyshko96}, \cite{Breit:Schill:Mlyn97}, \cite{LvRay09}%
, \cite{Scully:Zubairy97}, \cite{Klyshko98}, \cite{BachorRalph04},
\cite{Klyshko11}, \cite{Haroche13} and the references therein).

\section{The Wigner and Moyal Functions}

In coordinate representation, Wigner's functions can be derived by following
our analysis in the simplest case, when $\lambda=0$ \cite{KrySusVega13} (see
also \cite{Dod:Man:Man94}, \cite{HilletyetalWigner84}, \cite{Schleich01} and
\cite{Schradeetal95} for a general approach). For the multi-parameter
\textquotedblleft dynamical vacuum state\textquotedblright, when $n=0,$ for
example, the final result is given by%
\begin{equation}
W\left(  x,p,t\right)  =\frac{1}{\pi\sqrt{\omega}}\exp\left(  -Q\left(
U,V\right)  \right)  , \label{Wigner}%
\end{equation}
where%
\begin{equation}
Q\left(  U,V\right)  =\frac{\beta^{2}\left(  0\right)  \left[  \beta\left(
0\right)  U+\omega\varepsilon\left(  0\right)  \right]  ^{2}+\left[
2\alpha\left(  0\right)  U-\omega\left(  V-\delta\left(  0\right)  \right)
\right]  ^{2}}{\beta^{2}\left(  0\right)  \omega^{2}} \label{QuadraticWigner}%
\end{equation}
in the rotating $X=\omega x\cos\omega t-p\sin\omega t,$ $P=\omega x\sin\omega
t+p\sin\omega t$ and \textquotedblleft squeezing\textquotedblright%
\ coordinates%
\begin{equation}
U=\left\{
\begin{array}
[c]{c}%
\left[  \left(  X-P\right)  e^{\lambda t}+\left(  X+P\right)  e^{-\lambda
t}\right]  \medskip/2\ ,\\
Xe^{-\lambda t}%
\end{array}
\right.  \label{SqueezingU}%
\end{equation}%
\begin{equation}
V=\left\{
\begin{array}
[c]{c}%
\medskip\left[  \left(  P-X\right)  e^{\lambda t}+\left(  X+P\right)
e^{-\lambda t}\right]  /2\ ,\\
Pe^{\lambda t}%
\end{array}
\right.  \label{SqueezingV}%
\end{equation}
in the quantum phase space for the Hamiltonians (\ref{HamI}) and
(\ref{HamII}), respectively. Our result is consistent with \cite{Mollow67},
\cite{Agarwal87}, \cite{FernCollett88}, \cite{Dod:Man:Man94} and
\cite{Walls:Milburn} in the case of the initial vacuum state for the second
Hamiltonian. A similar consideration is valid for Moyal's functions
\cite{KrySusVega13}, \cite{Schleich01}. An example of generation of the
squeezed vacuum state is presented in Figure~1.
\begin{figure}[htbp]
\centering%
\scalebox{.75}%
{\includegraphics{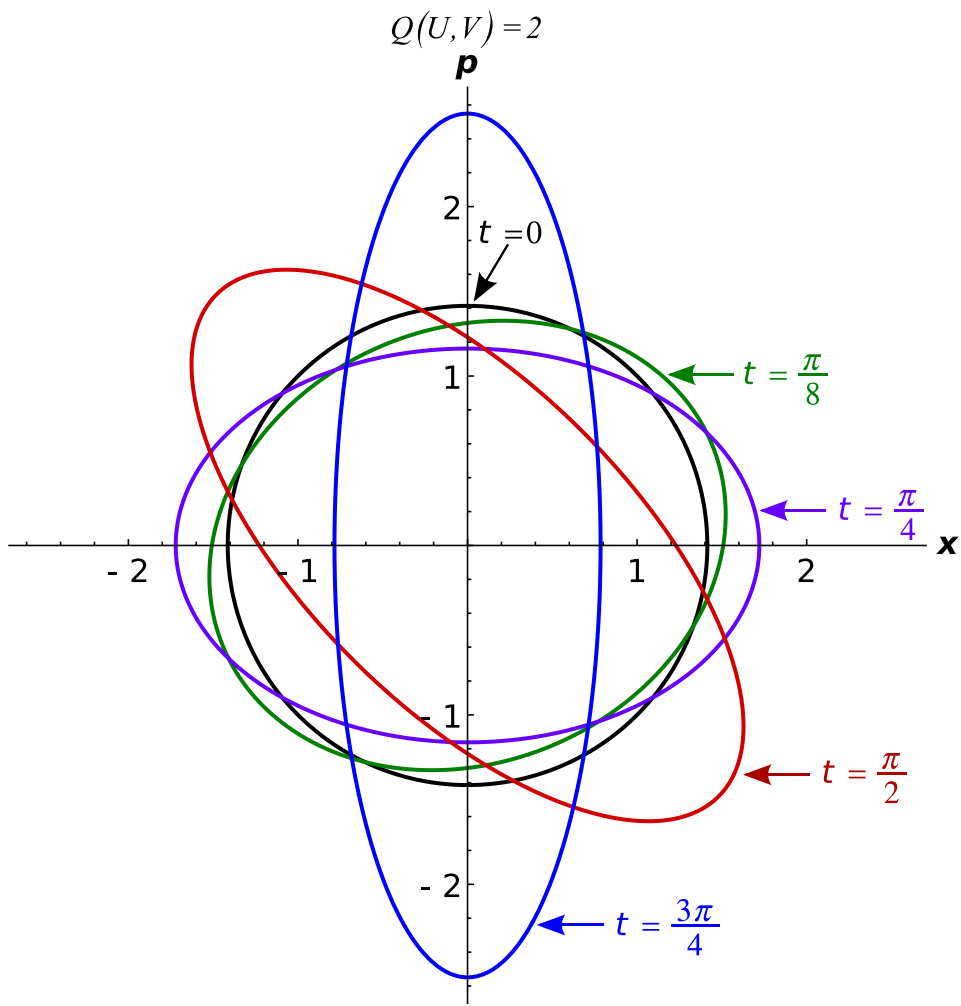}}
\caption{Phase space animation showing rotation and squeezing of contours
$Q\left(U, V\right)=2$ for the first Hamiltonian in (\ref{QuadraticWigner})--(\ref{SqueezingV})
with $\omega=1$ and $\lambda=0.25$ for the initial vacuum state.
The minimum-uncertainty squeezed states occur at $t_{\text{min}} = 0,\ \pi/4,\ 3\pi/4,$ etc.
(The color version of this figure is available only in the electronic edition.)}
\end{figure}

In summary, we have investigated, as explicitly as possible in coordinate
representation, all quantum statistical properties of the single mode
multi-parameter squeezed photon states in two classical models of optical
degenerate parametric oscillators. In principle, one can reproduce our
explicit time-dependent photon statistics in a more general operator approach
with the help of methods of representation theory (see, for example,
\cite{Marian91}, \cite{MarianMarianI93}, \cite{Schumaker86},
\cite{SchumakerCaves85} and Appendix~D). It will be of interest to everyone
who studies quantum optics and cavity QED. From the mathematical standpoint,
our results motivate further investigations of the Hamiltonian (\ref{RaiHam})
and the corresponding generalized Ince equation (\ref{RaiCharMu}). Last but
not least, it is worth noting that variable linear terms in creation and
annihilation operators, which correspond to a classical current
\cite{Glauber63}, \cite{Hanamuraetal07}, \cite{Scully:Zubairy97},
\cite{Walls:Milburn}, can be easily incorporated into this Hamiltonian in our
approach. Lossy medium models may also be taken into consideration. \medskip

\noindent\textbf{Acknowledgments.\/} We would like to thank Albert Boggess for
help and encouragement. This research was partially supported by AFOSR grant
FA9550-11-1-0220. The first named author was partially supported by
MICIIN/FEDER MTM2009--06973, gencat 2009SGR859 and DIDI -- Universidad del
Norte. One of the authors (ES) was also supported by the AMS-Simons Travel
Grants, with support provided by the Simons Foundation. Sergei Suslov thanks
Marlan O. Scully, Wolfgang Schleich and M. Suhail Zubairy for valuable
discussions during his visit to Institute for Quantum Science and Engineering,
Texas A\&M University. We are grateful to Victor V. Dodonov, Vladimir I.
Man'ko, Geza Giedke, Boris A. Malomed, Paulina Marian, Giuseppe Ruoso,
Christoph Koutschan, Jos\'{e} M. Vega-Guzm\'{a}n and Andreas Ruffing for
valuable comments and important references and to Kamal Barley for graphics enhancement.

\appendix

\section{Explicit Solutions of Ermakov-type System, Means and Variances}

The general solution of Ermakov-type system for the first Hamiltonian
(\ref{HamI}) can be found as follows%
\begin{align}
\alpha\left(  t\right)   &  =\frac{\omega}{2L\left(  t\right)  }\left[
\left(  4\alpha^{2}\left(  0\right)  +\beta^{4}\left(  0\right)  -\omega
^{2}\right)  \sin2\omega t\right. \label{AlphaI}\\
&  \left.  +\left(  4\alpha\left(  0\right)  \omega\cosh2\lambda t+\left(
4\alpha^{2}\left(  0\right)  +\beta^{4}\left(  0\right)  +\omega^{2}\right)
\sinh2\lambda t\right)  \cos2\omega t\right]  ,\nonumber\\
\beta\left(  t\right)   &  =\beta\left(  0\right)  \omega\sqrt{\frac
{2}{L\left(  t\right)  }},\label{BetaI}\\
\gamma\left(  t\right)   &  =-\frac{1}{2}\arctan\left(  \frac{\beta^{2}\left(
0\right)  \left(  \sinh\lambda t\ \cos\omega t+\cosh\lambda t\ \sin\omega
t\right)  }{M\left(  t\right)  }\right)  \label{GammaI}%
\end{align}
and%
\begin{align}
\delta\left(  t\right)   &  =\frac{2\omega}{L\left(  t\right)  }\left[
\left(  \delta\left(  0\right)  \omega\cos\omega t+\left(  2\alpha\left(
0\right)  \delta\left(  0\right)  +\beta^{3}\left(  0\right)  \varepsilon
\left(  0\right)  \right)  \sin\omega t\right)  \cosh\lambda t\right.
\label{DeltaI}\\
&  +\left.  \left(  \left(  2\alpha\left(  0\right)  \delta\left(  0\right)
+\beta^{3}\left(  0\right)  \varepsilon\left(  0\right)  \right)  \cos\omega
t+\delta\left(  0\right)  \omega\sin\omega t\right)  \sinh\lambda t\right]
,\nonumber\\
\varepsilon\left(  t\right)   &  =\sqrt{\frac{2}{L\left(  t\right)  }}\left[
\left(  \varepsilon\left(  0\right)  \omega\cos\omega t+\left(  2\alpha\left(
0\right)  \varepsilon\left(  0\right)  -\beta\left(  0\right)  \delta\left(
0\right)  \right)  \sin\omega t\right)  \cosh\lambda t\right.
\label{EpsilonI}\\
&  +\left.  \left(  \left(  2\alpha\left(  0\right)  \varepsilon\left(
0\right)  -\beta\left(  0\right)  \delta\left(  0\right)  \right)  \cos\omega
t+\varepsilon\left(  0\right)  \omega\sin\omega t\right)  \sinh\lambda
t\right]  ,\nonumber\\
\kappa\left(  t\right)   &  =\frac{1}{2L\left(  t\right)  }\left[  2\left(
\beta^{3}\left(  0\right)  \delta\left(  0\right)  \varepsilon\left(
0\right)  +\alpha\left(  0\right)  \left(  \delta^{2}\left(  0\right)
-\beta^{2}\left(  0\right)  \varepsilon^{2}\left(  0\right)  \right)  \right)
\cos2\omega t\right. \label{KappaI}\\
&  -\left(  2\left(  \beta^{3}\left(  0\right)  \delta\left(  0\right)
\varepsilon\left(  0\right)  +\alpha\left(  0\right)  \left(  \delta
^{2}\left(  0\right)  -\beta^{2}\left(  0\right)  \varepsilon^{2}\left(
0\right)  \right)  \right)  \right. \nonumber\\
&  \left.  +\left(  \left(  \delta^{2}\left(  0\right)  -\beta^{2}\left(
0\right)  \varepsilon^{2}\left(  0\right)  \right)  \omega\sin2\omega
t\right.  \right)  \cosh2\lambda t\nonumber\\
&  -\left(  \left(  \left(  \delta^{2}\left(  0\right)  -\beta^{2}\left(
0\right)  \varepsilon^{2}\left(  0\right)  \right)  \right)  \omega\right.
\nonumber\\
&  +\left.  2\left(  \beta^{3}\left(  0\right)  \delta\left(  0\right)
\varepsilon\left(  0\right)  +\alpha\left(  0\right)  \left(  \delta
^{2}\left(  0\right)  -\beta^{2}\left(  0\right)  \varepsilon^{2}\left(
0\right)  \right)  )\sin2\omega t\right)  \sinh2\lambda t\right]  ,\nonumber
\end{align}
where%
\begin{align}
L\left(  t\right)   &  =\left(  4\alpha^{2}\left(  0\right)  +\beta^{4}\left(
0\right)  +\omega^{2}+4\alpha\left(  0\right)  \omega\sin2\omega t\right)
\cosh2\lambda t\label{L(t)}\\
&  +\left(  4\alpha\left(  0\right)  \omega+\left(  4\alpha^{2}\left(
0\right)  +\beta^{4}\left(  0\right)  +\omega^{2}\right)  \sin2\omega
t\right)  \sinh2\lambda t\nonumber\\
&  -\left(  4\alpha^{2}\left(  0\right)  +\beta^{4}\left(  0\right)
-\omega^{2}\right)  \cos2\omega t\nonumber
\end{align}
and%
\begin{align}
M\left(  t\right)   &  =\left(  \omega\cos\omega t+2\alpha\left(  0\right)
\sin\omega t\right)  \cosh\lambda t\label{M(t)}\\
&  +\left(  2\alpha\left(  0\right)  \cos\omega t+\omega\sin\omega t\right)
\sinh\lambda t.\nonumber
\end{align}

For the second Hamiltonian (\ref{HamII}), in a similar fashion,%
\begin{align}
\alpha\left(  t\right)   &  =\omega\frac{\alpha\left(  0\right)  \omega
e^{2\lambda t}\cos2\omega t+\sin2\omega t\ \left(  \beta^{4}\left(  0\right)
+4\alpha^{2}\left(  0\right)  -\omega^{2}e^{4\lambda t}\right)  /4}{\beta
^{4}\left(  0\right)  \sin^{2}\omega t+\left(  2\alpha\left(  0\right)
\sin\omega t+\omega e^{2\lambda t}\cos\omega t\right)  ^{2}},\label{AlphaII}\\
\beta\left(  t\right)   &  =\frac{\omega\beta\left(  0\right)  e^{\lambda t}%
}{\sqrt{\beta^{4}\left(  0\right)  \sin^{2}\omega t+\left(  2\alpha\left(
0\right)  \sin\omega t+\omega e^{2\lambda t}\cos\omega t\right)  ^{2}}%
},\label{BetaII}\\
\gamma\left(  t\right)   &  =-\frac{1}{2}\arctan\frac{\beta^{2}\left(
0\right)  }{2\alpha\left(  0\right)  +\omega e^{2\lambda t}\cot\omega t}
\label{GammaII}%
\end{align}
and%
\begin{align}
\delta\left(  t\right)   &  =\omega e^{\lambda t}\frac{\varepsilon\left(
0\right)  \beta^{3}\left(  0\right)  \sin\omega t+\left(  2\alpha\left(
0\right)  \sin\omega t+\omega e^{2\lambda t}\cos\omega t\right)  \delta\left(
0\right)  }{\beta^{4}\left(  0\right)  \sin^{2}\omega t+\left(  2\alpha\left(
0\right)  \sin\omega t+\omega e^{2\lambda t}\cos\omega t\right)  ^{2}%
},\label{DeltaII}\\
\varepsilon\left(  t\right)   &  =\frac{\varepsilon\left(  0\right)  \left(
2\alpha\left(  0\right)  \sin\omega t+\omega e^{2\lambda t}\cos\omega
t\right)  -\beta\left(  0\right)  \delta\left(  0\right)  \sin\omega t}%
{\sqrt{\beta^{4}\left(  0\right)  \sin^{2}\omega t+\left(  2\alpha\left(
0\right)  \sin\omega t+\omega e^{2\lambda t}\cos\omega t\right)  ^{2}}},
\label{EpsilonII}%
\end{align}%
\begin{align}
\kappa\left(  t\right)   &  =\sin^{2}\omega t\frac{\varepsilon\left(
0\right)  \beta^{2}\left(  0\right)  \left(  \alpha\left(  0\right)
\varepsilon\left(  0\right)  -\beta\left(  0\right)  \delta\left(  0\right)
\right)  -\alpha\left(  0\right)  \delta^{2}\left(  0\right)  }{\beta
^{4}\left(  0\right)  \sin^{2}\omega t+\left(  2\alpha\left(  0\right)
\sin\omega t+\omega e^{2\lambda t}\cos\omega t\right)  ^{2}}\label{KappaII}\\
&  +\frac{\omega e^{2\lambda t}\sin2\omega t\left(  \varepsilon^{2}\left(
0\right)  \beta^{2}\left(  0\right)  -\delta^{2}\left(  0\right)  \right)
}{4\left(  \beta^{4}\left(  0\right)  \sin^{2}\omega t+\left(  2\alpha\left(
0\right)  \sin\omega t+\omega e^{2\lambda t}\cos\omega t\right)  ^{2}\right)
}.\nonumber
\end{align}
(In both cases, we assume without loss of generality that $\gamma\left(
0\right)  =\kappa\left(  0\right)  =0.)$

The means and variances are given by%
\begin{align}
\left\langle \left.  \widehat{q}\right.  \right\rangle  &  =\frac{\left(
2\alpha\left(  0\right)  \varepsilon\left(  0\right)  -\beta\left(  0\right)
\delta\left(  0\right)  \right)  \sin\omega t-\varepsilon\left(  0\right)
\omega\cos\omega t}{\beta\left(  0\right)  \omega}\cosh\lambda t\label{XI}\\
&  +\frac{\left(  2\alpha\left(  0\right)  \varepsilon\left(  0\right)
-\beta\left(  0\right)  \delta\left(  0\right)  \right)  \cos\omega
t-\varepsilon\left(  0\right)  \omega\sin\omega t}{\beta\left(  0\right)
\omega}\sinh\lambda t,\nonumber\\
\left\langle \left.  \widehat{p}\right.  \right\rangle  &  =\frac
{\varepsilon\left(  0\right)  \omega\sin\omega t-\left(  2\alpha\left(
0\right)  \varepsilon\left(  0\right)  -\beta\left(  0\right)  \delta\left(
0\right)  \right)  \cos\omega t}{\beta\left(  0\right)  \omega}\cosh\lambda
t\label{PI}\\
&  -\frac{\left(  2\alpha\left(  0\right)  \varepsilon\left(  0\right)
-\beta\left(  0\right)  \delta\left(  0\right)  \right)  \sin\omega
t-\varepsilon\left(  0\right)  \omega\cos\omega t}{\beta\left(  0\right)
\omega}\sinh\lambda t,\nonumber
\end{align}%
\begin{align}
\sigma_{q}\left(  t\right)   &  =\frac{4\alpha^{2}\left(  0\right)  +\beta
^{4}\left(  0\right)  +\omega^{2}+4\alpha\left(  0\right)  \omega\sin2\omega
t}{4\beta^{2}\left(  0\right)  \omega^{2}}\cosh2\lambda t\label{SigmaXI}\\
&  +\frac{4\alpha\left(  0\right)  \omega+\left(  4\alpha^{2}\left(  0\right)
+\beta^{4}\left(  0\right)  +\omega^{2}\right)  \sin2\omega t}{4\beta
^{2}\left(  0\right)  \omega^{2}}\sinh2\lambda t\nonumber\\
&  -\frac{4\alpha^{2}\left(  0\right)  +\beta^{4}\left(  0\right)  -\omega
^{2}}{4\beta^{2}\left(  0\right)  \omega^{2}}\cos2\omega t,\nonumber\\
\sigma_{p}\left(  t\right)   &  =\frac{4\alpha^{2}\left(  0\right)  +\beta
^{4}\left(  0\right)  +\omega^{2}-4\alpha\left(  0\right)  \omega\sin2\omega
t}{4\beta^{2}\left(  0\right)  }\cosh2\lambda t\label{SigmaPI}\\
&  +\frac{4\alpha\left(  0\right)  \omega-\left(  4\alpha^{2}\left(  0\right)
+\beta^{4}\left(  0\right)  +\omega^{2}\right)  \sin2\omega t}{4\beta
^{2}\left(  0\right)  }\sinh2\lambda t\nonumber\\
&  +\frac{4\alpha^{2}\left(  0\right)  +\beta^{4}\left(  0\right)  -\omega
^{2}}{4\beta^{2}\left(  0\right)  }\cos2\omega t\nonumber
\end{align}
and%
\begin{align}
\left\langle \left.  \widehat{q}\right.  \right\rangle  &  =-\frac
{\varepsilon\left(  0\right)  }{\beta\left(  0\right)  }e^{\lambda t}%
\cos\omega t+\frac{1}{\omega}\left(  \delta\left(  0\right)  -\frac
{2\alpha\left(  0\right)  \varepsilon\left(  0\right)  }{\beta\left(
0\right)  }\right)  e^{-\lambda t}\sin\omega t,\label{XII}\\
\left\langle \left.  \widehat{p}\right.  \right\rangle  &  =\left(
\delta\left(  0\right)  -\frac{2\alpha\left(  0\right)  \varepsilon\left(
0\right)  }{\beta\left(  0\right)  }\right)  e^{-\lambda t}\cos\omega
t+\omega\frac{\varepsilon\left(  0\right)  }{\beta\left(  0\right)
}e^{\lambda t}\sin\omega t, \label{PII}%
\end{align}%
\begin{align}
\sigma_{q}\left(  t\right)   &  =\frac{\left(  4\alpha^{2}\left(  0\right)
+\beta^{4}\left(  0\right)  \right)  e^{-2\lambda t}+\omega^{2}e^{2\lambda t}%
}{4\beta^{2}\left(  0\right)  \omega^{2}}+\frac{\alpha\left(  0\right)
}{\beta^{2}\left(  0\right)  \omega}\sin2\omega t\label{SigmaXII}\\
&  -\frac{\left(  4\alpha^{2}\left(  0\right)  +\beta^{4}\left(  0\right)
\right)  e^{-2\lambda t}-\omega^{2}e^{2\lambda t}}{4\beta^{2}\left(  0\right)
\omega^{2}}\cos2\omega t,\nonumber\\
\sigma_{p}\left(  t\right)   &  =\frac{\left(  4\alpha^{2}\left(  0\right)
+\beta^{4}\left(  0\right)  \right)  e^{-2\lambda t}+\omega^{2}e^{2\lambda t}%
}{4\beta^{2}\left(  0\right)  }-\frac{\alpha\left(  0\right)  \omega}%
{\beta^{2}\left(  0\right)  }\sin2\omega t\label{SigmaPII}\\
&  +\frac{\left(  4\alpha^{2}\left(  0\right)  +\beta^{4}\left(  0\right)
\right)  e^{-2\lambda t}-\omega^{2}e^{2\lambda t}}{4\beta^{2}\left(  0\right)
}\cos2\omega t\nonumber
\end{align}
for the Hamiltonians (\ref{HamI}) and (\ref{HamII}), respectively.

In this article, we take the initial data of the corresponding Ermakov-type
systems as constants of motion (that naturally describes multi-parameter
squeezing). In coordinate representation, their physical meaning is clear from
the explicit wave function (\ref{WaveFunctionN}) when $t=0.$ Another useful
set of integrals of motion consists of the expectation values and variances at
$t=0:$ $\left.  \left\langle \left.  \widehat{q}\right.  \right\rangle
\right\vert _{t=0},$ $\left.  \left\langle \left.  \widehat{p}\right.
\right\rangle \right\vert _{t=0},$ $\left.  \sigma_{q}\left(  0\right)
\right\vert _{n=0},$ $\left.  \sigma_{p}\left(  0\right)  \right\vert _{n=0}$
and $\left.  \sigma_{pq}\left(  0\right)  \right\vert _{n=0};$ the latter are
related by (\ref{Dinvariant}) (see Ref.~\cite{Dod:Man:Man94} for more details).

\section{Hamiltonian Expansion}

The single mode Hamiltonian, $\widehat{H}=\left(  \left.  \widehat{p}\right.
^{2}+\omega^{2}\left.  \widehat{q}\right.  ^{2}\right)  /2,$ can be rewritten
as follows
\begin{align}
\widehat{H}  &  =\left(  \frac{4\alpha^{2}-\beta^{4}+\omega^{2}}{4\beta^{2}%
}-i\alpha\right)  \left.  \widehat{a}\left(  t\right)  \right.  ^{2}+\left(
\frac{4\alpha^{2}-\beta^{4}+\omega^{2}}{4\beta^{2}}+i\alpha\right)  \left.
\widehat{a}^{\dagger}\left(  t\right)  \right.  ^{2}\label{HAB}\\
&  +\frac{4\alpha^{2}+\beta^{4}+\omega^{2}}{4\beta^{2}}\left[  \widehat{a}%
\left(  t\right)  \widehat{a}^{\dagger}\left(  t\right)  +\widehat{a}%
^{\dagger}\left(  t\right)  \widehat{a}\left(  t\right)  \right] \nonumber\\
&  +\sqrt{2}\left[  \frac{\alpha}{\beta}\left(  \delta-\frac{2\alpha
\varepsilon}{\beta}\right)  -\frac{\varepsilon\omega^{2}}{2\beta^{2}}%
-\frac{i\beta}{2}\left(  \delta-\frac{2\alpha\varepsilon}{\beta}\right)
\right]  \widehat{a}\left(  t\right) \nonumber\\
&  +\sqrt{2}\left[  \frac{\alpha}{\beta}\left(  \delta-\frac{2\alpha
\varepsilon}{\beta}\right)  -\frac{\varepsilon\omega^{2}}{2\beta^{2}}%
+\frac{i\beta}{2}\left(  \delta-\frac{2\alpha\varepsilon}{\beta}\right)
\right]  \widehat{a}^{\dagger}\left(  t\right) \nonumber\\
&  +\frac{1}{2}\left(  \delta-\frac{2\alpha\varepsilon}{\beta}\right)
^{2}+\frac{\varepsilon^{2}\omega^{2}}{2\beta^{2}}\nonumber
\end{align}
in terms of variable creation and annihilation operators $\widehat{a}\left(
t\right)  =e^{2i\gamma}\widehat{b}\left(  t\right)  $ and $\widehat{a}%
^{\dagger}\left(  t\right)  =e^{-2i\gamma}\widehat{b}^{\dagger}\left(
t\right)  $ which is convenient for evaluation of the corresponding matrix
elements in section$~$5. More details can be found in \cite{KretalSus13},
\cite{KrySusVega13}, \cite{SanSusVin}.

\section{Interaction Picture}

Formally, from (\ref{Schroedinger})--(\ref{RaiHam}) one gets,%
\begin{equation}
\psi=e^{-i\left(  \widehat{a}^{\dagger}\ \widehat{a}+1/2\right)  \omega
t}\left.  \chi\right.  ,\qquad i\chi_{t}=\widehat{V}\chi, \label{FormalSub}%
\end{equation}
where%
\begin{align}
\widehat{V}  &  =-\frac{\lambda\left(  t\right)  }{2}e^{i\left(
\widehat{a}^{\dagger}\ \widehat{a}\right)  \omega t}\left(  e^{i\left(
2\omega t+\phi\left(  t\right)  \right)  }\ \left.  \widehat{a}\right.
^{2}+e^{-i\left(  2\omega t+\phi\left(  t\right)  \right)  }\left(
\widehat{a}^{\dagger}\right)  ^{2}\right)  e^{-i\left(  \widehat{a}^{\dagger
}\ \widehat{a}\right)  \omega t}\nonumber\\
&  =-\frac{\lambda\left(  t\right)  }{2}\left(  e^{i\phi\left(  t\right)
}\ \left.  \widehat{a}\right.  ^{2}+e^{-i\phi\left(  t\right)  }\left(
\widehat{a}^{\dagger}\right)  ^{2}\right)  \label{InterHam}%
\end{align}
with the help of a familiar formal identity%
\begin{equation}
e^{i\left(  \widehat{a}^{\dagger}\ \widehat{a}\right)  \omega t}\left.
\widehat{a}\right.  e^{-i\left(  \widehat{a}^{\dagger}\ \widehat{a}\right)
\omega t}=\left.  \widehat{a}\right.  e^{-i\omega t} \label{InterIdentity}%
\end{equation}
and its conjugate. As a result, in interaction picture:%
\begin{equation}
\widehat{V}=\left\{
\begin{array}
[c]{c}%
-\lambda\left(  \left.  \widehat{a}\right.  ^{2}+\left(  \widehat{a}^{\dagger
}\right)  ^{2}\right)  /2=\lambda\left(  \left.  \widehat{p}\right.
^{2}-\omega^{2}\left.  \widehat{q}\right.  ^{2}\right)  /\left(
2\omega\right)  \ ,\\
\lambda\left(  \left.  \widehat{a}\right.  ^{2}-\left(  \widehat{a}^{\dagger
}\right)  ^{2}\right)  /\left(  2i\right)  =\lambda\left(  \left.
\widehat{p}\right.  \left.  \widehat{q}\right.  +\left.  \widehat{q}\right.
\left.  \widehat{p}\right.  \right)  /2\quad\quad
\end{array}
\right.  \label{InterHams}%
\end{equation}
for the original Hamiltonians (\ref{HamI}) and (\ref{HamII}), respectively.

In coordinate representation, the formal substitution (\ref{FormalSub}) takes
the form
\begin{equation}
\psi\left(  x,t\right)  =\int_{-\infty}^{\infty}G_{\omega}\left(
x,y,t\right)  \ \chi\left(  y,t\right)  \ dy \label{RealSubst}%
\end{equation}
in $\mathcal{L}^{2}\left(
\mathbb{R}
\right)  ,$ where%
\begin{equation}
G_{\omega}\left(  x,y,t\right)  =\sqrt{\frac{\omega}{2\pi i\sin\omega t}%
}\ \exp\left(  i\omega\frac{\left(  x^{2}+y^{2}\right)  \cos\omega
t-2xy}{2\sin\omega t}\right)  \label{GreenFunctionOsc}%
\end{equation}
is a familiar Green's function for quantum harmonic oscillator.

In interaction picture, solutions of the corresponding Cauchy initial value
problems for wave function $\chi\left(  x,t\right)  $ can be found as follows%
\begin{equation}
i\chi_{t}+\frac{\lambda}{2\omega}\left(  \chi_{xx}+\omega^{2}x^{2}\chi\right)
=0,\qquad\chi\left(  x,t\right)  =\int_{-\infty}^{\infty}G_{\lambda}\left(
x,y,t\right)  \ \psi_{0}\left(  y\right)  \ dy, \label{CIVPI}%
\end{equation}
where%
\begin{equation}
G_{\lambda}\left(  x,y,t\right)  =\sqrt{\frac{\omega}{2\pi i\sinh\lambda t}%
}\ \exp\left(  i\omega\frac{\left(  x^{2}+y^{2}\right)  \cosh\lambda
t-2xy}{2\sinh\lambda t}\right)  , \label{GreenFunctionUnOsc}%
\end{equation}
and%
\begin{equation}
\chi_{t}+\lambda x\chi_{x}+\frac{\lambda}{2}\chi=0,\qquad\chi\left(
x,t\right)  =e^{-\lambda t/2}\psi_{0}\left(  xe^{-\lambda t}\right)  ,
\label{CIVPII}%
\end{equation}
in the first case and the second cases, respectively. This consideration gives
an independent verification of our Green's functions%
\begin{equation}
G\left(  x,y,t\right)  =\frac{e^{i\left(  \alpha_{0}x^{2}+\beta_{0}%
xy+\gamma_{0}y^{2}\right)  }}{\sqrt{2\pi i\mu_{0}}}=\left\{
\begin{array}
[c]{c}%
{\displaystyle\int_{-\infty}^{\infty}}
G_{\omega}\left(  x,z,t\right)  G_{\lambda}\left(  z,y,t\right)
\ dz,\bigskip\\
e^{\lambda t/2}\ G_{\omega}\left(  x,ye^{\lambda t},t\right)  \qquad\qquad
\end{array}
\right.  \label{GreenOL}%
\end{equation}
for the Hamiltonians (\ref{HamI}) and (\ref{HamII}), respectively; cf.
Eqs.~(\ref{AlphaI-II})--(\ref{GammaI-II}).

\section{Canonical Transformation}

Two familiar identities are valid:%
\begin{align}
e^{\xi^{\ast}\left.  \widehat{a}\right.  -\xi\left.  \widehat{a}^{\dagger
}\right.  }\left.  \widehat{a}\right.  e^{-\xi^{\ast}\left.  \widehat{a}%
\right.  +\xi\left.  \widehat{a}^{\dagger}\right.  } &  =\left.
\widehat{a}\right.  +\xi,\label{OperIds}\\
e^{\left(  e^{2i\varphi}\left.  \widehat{a}\right.  ^{2}-e^{-2i\varphi}\left(
\widehat{a}^{\dagger}\right)  ^{2}\right)  \tau/2}\left.  \widehat{a}\right.
e^{-\left(  e^{2i\varphi}\left.  \widehat{a}\right.  ^{2}-e^{-2i\varphi
}\left(  \widehat{a}^{\dagger}\right)  ^{2}\right)  \tau/2} &  =\left(
\cosh\tau\right)  \left.  \widehat{a}\right.  +e^{-2i\varphi}\left(  \sinh
\tau\right)  \left.  \widehat{a}^{\dagger}\right.  \label{OpIdSqueeze}%
\end{align}
formally for the single-mode displacement \cite{Glauber63} and squeeze
\cite{Stoler70}, \cite{Stoler71} operators, respectively (see, for example,
\cite{Schumaker86}, \cite{SchumakerCaves85} for more details). With the aid of
(\ref{InterIdentity}) and (\ref{OperIds})--(\ref{OpIdSqueeze}), one can verify
that the canonical transformation (\ref{LinDynInv}), namely,%
\begin{equation}
\widehat{b}\left(  t\right)  =\mathbf{U}\left(  t\right)  \left.
\widehat{a}\right.  \mathbf{U}^{-1}\left(  t\right)  =\frac{e^{-2i\gamma}%
}{\sqrt{2}}\left(  \beta\widehat{q}+\varepsilon+i\frac{\widehat{p}%
-2\alpha\widehat{q}-\delta}{\beta}\right)  \label{BCanonical}%
\end{equation}
holds for the unitary operator of the form:%
\begin{equation}
\mathbf{U}\left(  t\right)  =e^{i\left(  \widehat{a}^{\dagger}\ \widehat{a}%
\right)  \theta}e^{\left(  e^{2i\varphi}\left.  \widehat{a}\right.
^{2}-e^{-2i\varphi}\left(  \widehat{a}^{\dagger}\right)  ^{2}\right)  \tau
/2}e^{\xi^{\ast}\left.  \widehat{a}\right.  -\xi\left.  \widehat{a}^{\dagger
}\right.  }e^{2i\left(  \widehat{a}^{\dagger}\ \widehat{a}\right)  \gamma
}\label{UofT}%
\end{equation}
provided%
\begin{align}
&  \frac{1}{\sqrt{\omega}}\left(  \beta-\frac{2i\alpha}{\beta}\right)
+\frac{\sqrt{\omega}}{\beta}=2e^{-i\theta}\cosh\tau,\label{ABTwoParameters}\\
&  \frac{1}{\sqrt{\omega}}\left(  \beta-\frac{2i\alpha}{\beta}\right)
-\frac{\sqrt{\omega}}{\beta}=2e^{i\left(  \theta-2\varphi\right)  }\sinh
\tau,\label{ABTParametersPhy}%
\end{align}
and%
\begin{equation}
\xi\sqrt{2}=\varepsilon-i\frac{\delta}{\beta}=-i\left(  \frac{\delta\left(
0\right)  }{\beta\left(  0\right)  }+i\varepsilon\left(  0\right)  \right)
e^{2i\gamma}.\label{CShift}%
\end{equation}
Therefore, the time-dependent parameters of our single-mode \textquotedblleft
multi-parameter squeeze/evolution operator\textquotedblright\ (\ref{UofT}),
namely, $\theta(t),$ $\tau(t),$ $\varphi(t)$ and $\xi(t),$ are determined in
terms of solutions of the corresponding Ermakov-type system as follows%
\begin{align}
&  \tan\theta\left(  t\right)  =\frac{2\alpha}{\beta^{2}+\omega},\quad
\tan2\varphi\left(  t\right)  =\frac{-4\alpha\beta^{2}}{4\alpha^{2}+\omega
^{2}-\beta^{4}},\label{ParametersThetaTau}\\
&  4\left[  \cosh\tau\left(  t\right)  \right]  ^{2}=\left(  \frac{\beta
}{\sqrt{\omega}}+\frac{\sqrt{\omega}}{\beta}\right)  ^{2}+\frac{4\alpha^{2}%
}{\omega\beta^{2}},\label{ParameterTau}\\
&  4\left[  \sinh\tau\left(  t\right)  \right]  ^{2}=\left(  \frac{\beta
}{\sqrt{\omega}}-\frac{\sqrt{\omega}}{\beta}\right)  ^{2}+\frac{4\alpha^{2}%
}{\omega\beta^{2}}\label{ParameterTauSh}%
\end{align}
(see also (\ref{CShift})).

Assuming that the vacuum state is nondegenerate, one gets $\left.
\widehat{a}\right.  \left(  \mathbf{U}^{-1}\left(  t\right)  \left\vert
\psi_{0}\left(  t\right)  \right\rangle \right)  =0$ and $\left\vert \psi
_{0}\left(  t\right)  \right\rangle =\mathbf{U}\left(  t\right)  \left\vert
0\right\rangle .$ As a result,
\begin{equation}
\left\vert \psi_{n}\left(  t\right)  \right\rangle =\frac{1}{\sqrt{n!}}\left(
\widehat{b}^{\dagger}\left(  t\right)  \right)  ^{n}\left\vert \psi_{0}\left(
t\right)  \right\rangle =\mathbf{U}\left(  t\right)  \left(  \frac{1}%
{\sqrt{n!}}\left(  \widehat{a}^{\dagger}\right)  ^{n}\left\vert 0\right\rangle
\right)  =\mathbf{U}\left(  t\right)  \left\vert n\right\rangle
\label{TimeSqueezeOper}%
\end{equation}
in terms of standard Fock's number states. Our evolution operator
$\mathbf{U}\left(  t\right)  $ satisfies, formally, the time-dependent
Schr\"{o}dinger equation (\ref{Schroedinger}). In addition, one may conclude
that the minimum-uncertainty squeezed states occur when $\alpha\left(
t_{\text{min}}\right)  =0$ and $n=0$ (see, for example, Eq.~(5.5) of
Ref.~\cite{KretalSus13}). Indeed, only at these moments of time, by
(\ref{ParametersThetaTau}) the following identities hold, $\theta\left(
t_{\text{min}}\right)  =\varphi\left(  t_{\text{min}}\right)  =\alpha\left(
t_{\text{min}}\right)  =0,$ and a traditional definition of single-mode
squeeze operator from Refs.~\cite{Schumaker86}, \cite{SchumakerCaves85},
\cite{Stoler70}, \cite{Stoler71} can be used, say, \textquotedblleft
stroboscopically\textquotedblright. In general, the variable squeeze operator
in (\ref{UofT}) and (\ref{TimeSqueezeOper}) should be applied for our initial
\textquotedblleft multi-parameter squeezed number states\textquotedblright%
\ given by $\left\vert \psi_{n}\left(  0\right)  \right\rangle =\mathbf{U}%
\left(  0\right)  \left\vert n\right\rangle ;$ see also \cite{Marhic78},
\cite{LopSusVegaHarm}, \cite{Lop:Sus:VegaGroup}, \cite{KrySusVega13}. (The
traditional squeeze operator corresponds to a special case $\alpha\left(
0\right)  =\theta\left(  0\right)  =\varphi\left(  0\right)  =0$ and
$\tau\left(  0\right)  =\left(  1/2\right)  \ln\left(  \beta^{2}\left(
0\right)  /\omega\right)  .)$ Last but not least, this consideration allows
one to re-evaluate the photon amplitudes (\ref{CoeffFinalDependent}) in a pure
algebraic fashion similar to Refs.~\cite{VenkSatya85}, \cite{Marian91},
\cite{Marian92}.

In this article, we use the Schr\"{o}dinger picture which is more convenient
in the theory of quantum measurement\footnote{A detailed analysis of different
representations of quantum mechanics and quantum optics is given in
Refs.~\cite{Sudbery86}, \cite{Klyshko94}, \cite{Klyshko96}, \cite{Klyshko98}%
.}. In the Heisenberg picture, one gets%
\begin{equation}
\widehat{q}\left(  t\right)  =\mathbf{U}^{-1}\left(  t\right)  \left.
\widehat{q}\right.  \mathbf{U}\left(  t\right)  =\frac{1}{\beta\sqrt{2}%
}\left(  e^{2i\gamma}\left.  \widehat{a}\right.  +\left.  \widehat{a}%
^{\dagger}\right.  e^{-2i\gamma}\right)  -\frac{\varepsilon}{\beta
}\label{HeisenbergQ}%
\end{equation}
and%
\begin{align}
\widehat{p}\left(  t\right)   &  =\mathbf{U}^{-1}\left(  t\right)  \left.
\widehat{p}\right.  \mathbf{U}\left(  t\right)  =\frac{\alpha\sqrt{2}}{\beta
}\left(  e^{2i\gamma}\left.  \widehat{a}\right.  +\left.  \widehat{a}%
^{\dagger}\right.  e^{-2i\gamma}\right)  \label{HeisenbergP}\\
&  +\frac{\beta}{i\sqrt{2}}\left(  e^{2i\gamma}\left.  \widehat{a}\right.
-\left.  \widehat{a}^{\dagger}\right.  e^{-2i\gamma}\right)  +\delta
-\frac{2\alpha\varepsilon}{\beta}\nonumber
\end{align}
with the aid of (\ref{BCanonical}). The standard equations of motion hold,%
\begin{equation}
i\frac{d}{dt}\widehat{p}\left(  t\right)  =\left[  \widehat{p}\left(
t\right)  ,\ \widehat{\mathcal{H}}\left(  t\right)  \right]  ,\qquad i\frac
{d}{dt}\widehat{q}\left(  t\right)  =\left[  \widehat{q}\left(  t\right)
,\ \widehat{\mathcal{H}}\left(  t\right)  \right]  ,\label{HeisenbergPQ}%
\end{equation}
where $\ \widehat{\mathcal{H}}\left(  t\right)  =\mathbf{U}^{-1}\left(
t\right)  \widehat{H}\left(  t\right)  \mathbf{U}\left(  t\right)  .$ All
information about the state of radiation field is now encoded into the
operators (\ref{HeisenbergQ})--(\ref{HeisenbergP}) in the form of initial
data/constants of motion of the corresponding Ermakov-type system. The field
evolution is also completely determined in terms of explicit solutions of this
system. The expectation values of operators $\widehat{q}\left(  t\right)  $
and $\widehat{p}\left(  t\right)  $ with respect to Fock's states coincide, of
course, with those found in Ref.~\cite{KretalSus13} in the Schr\"{o}dinger picture.

\end{document}